\documentclass[aps,prb,showpacs]{revtex4}
\usepackage{graphicx}

\begin{document}

\title{Theory of single atom manipulation with a scanning probe tip: force signatures, constant-height and 
constant-force scans}

\author{\surname{Laurent} Pizzagalli\footnote{Author to whom correspondence should be sent. Phone: 33 549 496 833. Fax: 33 549 496 692. 
Email: Laurent.Pizzagalli@univ-poitiers.fr}}

\affiliation{Laboratoire de M\'etallurgie Physique, UMR 6630, Universit\'e de Poitiers, SP2MI, BP 30179, 
86962 Futuroscope Chasseneuil Cedex, France}

\author{\surname{Alexis} Baratoff}

\affiliation{Department of Physics and Astronomy, Basel University, 
Klingelbergstr. 82, CH-4056 BASEL,
Switzerland}

\begin{abstract}

We report theoretical results predicting the atomic manipulation of a silver atom on a Si(001) surface by a scanning probe tip, and 
providing several new insights about the manipulation 
phenomena. A molecular mechanics technique has been used, the system being described by a quantum chemistry method 
for the short-range interactions and 
an analytical model for the long-range ones. Taking into account several shapes, orientations, and chemical 
natures of the scanning tip, we observed four different ways to 
manipulate the deposited atom in a constant-height mode. In particular, the manipulation is predicted to be 
possible with a Si(111) tip for different 
tip shapes and adatom locations on the silicon surface. The calculation of the forces during the manipulation 
revealed that specific variations can be associated 
with each kind of process. These force signatures, like the tip height signatures observed in STM experiments, 
could be used to deduce the process involved in an 
experiment. Finally, we present preliminary results about the manipulation in constant-force mode. 

\end{abstract}

\maketitle

\section{Introduction}

Originally designed to image surfaces, scanning probe techniques\cite{Bin82PRL,Bin86PRL,Wic93ACAD} have also rapidly proved 
to be efficient for acting on the surfaces at the nanoscopic 
scale\cite{Bin99RMP}. First evidences came from the observation of the consequences of the 
contact between the scanning probe tip and the studied surface. For instance, Li et al observed that 
the mechanical interaction between their Scanning Tunneling 
Microscope (STM) tungsten tip and one Au surface led to the 
creation of nanoscopic pits\cite{Li89APL}. Similar indentation has also been performed on dichalcogenide WSe$_2$ 
surfaces \cite{Fuc90EPL,Sch92PSS}. 
Another kind of nano-patterning was achieved by the formation of mounds on the surface 
by field-emitting gold tip atoms\cite{Mam90PRL}. Dimensions of these created 
nanostructures were typically 100-300~\AA. However, 
manipulation with scanning probe techniques made a major step when 
state-of-the-art experiments performed in the group of D.~Eigler have shown  
that the STM tip could even control a unique deposited atom with an impressive accuracy \cite{Eig90NAT}. According to the 
notation of Stroscio and Eigler, manipulation processes can be classified 
in two categories\cite{Str91SCI}. {\it Vertical} processes involve 
atom transfers between the tip and the surface, voltage pulse or direct tip-surface 
contact being used for the atom transfers.
Adsorbates or native surface atoms can be 
either permanently extracted by the 
tip\cite{Lyo91SCI,Avo93NATO,Kob93NATO,Gre94PMB,Hos95JVST,She95SCI,Avo95ACR,Hua95JJAP,Sal96JVST,She97SS,Hua98JSA,Duj98PRL} 
or redeposited in another location\cite{Eig91NAT,Eig93NATO,Mey97SS,Mey01CMS}. This mode has been theoretically studied by several 
groups\cite{Joa92EPL,Gir92CP,Joa93NATO1,Lan93NATO,Tso95JJAP,Bul96PRB,Sto98PRL}. In the {\it lateral} category, manipulated objects 
always stay on the surface\cite{Str91SCI}. It is the scanning probe tip which either (i) trap the object in the 
surrounding of its apex and slide it\cite{Eig90NAT,Mey97SS,Zep92ULT,Bar97PRL}, 
(ii) push the object along the scan path\cite{Bet95APL,Jun95APL,Jun96SCI,Cub96APL,Cub98APA,Bau98NANO,Jay99PRB}, 
or (iii) control the diffusion by voltage pulse\cite{Whi91SCI,Bou98APA}.

It seems that the potential of lateral manipulation has not been 
completely explored. Successful experiments have been performed for very small objects, like Xe atoms 
on Ni(110)\cite{Eig90NAT}, CO molecules and Pt atoms on Pt(111)\cite{Zep92ULT},
Cu, Pb atoms, and CO, Pb$_2$, C$_2$H$_4$ molecules on Cu(211)\cite{Mey97SS,Bar97PRL,Mey99APA}. However, 
in these studies, all supporting substrates are closed metallic surfaces. Their low diffusion energy barriers make the 
manipulation easier. Theoretically, several 
calculations of the lateral manipulation of one atom have been 
done, and it has been found that single atoms can be manipulated on metallic (Pt(111)\cite{Bul96PRB}, 
Cu(110) \cite{Bou93PRB,Bou97PRB}, Cu(211) \cite{Kur99PRL})
or insulating (NaCl(001) \cite{Bou94PRB}) substrates. However, only one kind of tip is usually considered, which limits the 
scope of the results. Another limitation concerns the substrate. To our knowledge, the manipulation of 
one unique atom on a surface having a significant technological importance has not been achieved yet.
The Si(001) surface, heavily used in the electronic industry, is an interesting and still unprobed candidate.
In a previous paper, preliminary calculations had shown that 
one single Ag atom deposited on the Si(001) surface may be manipulated 
by a tip during a constant height scan \cite{Piz97SSL}. The complete results 
of the study are presented and discussed in this paper. First, we focus on 
constant height manipulation. In particular, we analyze the influence of several  
parameters, such as the shape of the tip, its orientation, and its chemical composition. 
Secondly, we show original results about the variation of the forces acting on the tip during different manipulation 
processes. These variations are then discussed in relation with the recent works on STM current 
signatures\cite{Bar97PRL,Bar97CPL,Bou99PRB}. Finally, in the last section, we discuss the possibility to perform 
constant-force manipulation scans with an Atomic Force Microscope (AFM) tip, and we show some examples. 

\section{Models}

The silicon (001) surface is now very well understood, thanks to several 
experimental\cite{Bad94JVST,Toc94PRB,Mun95PRL} and theoretical studies\cite{Cha79PRL,Rob90SS,Ram95PRB}. In its 
ground state, the surface is c(4$\times$2) reconstructed, with alternating asymmetric dimers. A 
(2$\times$1) reconstruction is observed at room temperature because the asymmetric dimers oscillate very rapidly 
and appear to be symmetric on the timescale of experiments. When a silver atom is deposed on the Si(001) surface, 
most of the experimental studies agreed that the adatom adsorb preferentially in the cave site, i.e. between two Si dimers 
belonging to two adjacent dimer rows\cite{Sam89PRL,Has90JVST,Lin93PRB1,Win94JVST,Kim95PRB}. On the Fig.~1, we have 
represented the adsorption energy map for one Ag atom deposited on the Si(001) surface\cite{Oko97SSL}. The cave site is 
the most stable, though its corresponding adsorption energy is very close to the energy of the bond site (Ag atom bonded to 
one dimerized Si atoms). This quasi degeneracy could explain why the Ag atom is sometimes found in 
the bond site\cite{Yak98JSA}. Here, most of the calculations have been carried out with the silver atom initially 
adsorbed in the cave site. However, we have also considered the case of adsorption in the bond site for selected examples. 

Two complementary models have been used to represent our system which is composed of the scanning probe tip, the 
Ag atom and the Si(001) 
surface. First, the short-range chemical forces between the very end of the tip, the adatom and the surface are described by 
the Atom Superposition and Electron Delocalization (ASED) method\cite{Nat90PRB}. This extension of the Extended H\"{u}ckel Molecular Orbitals 
(EHMO) model has been successfully used for the study of adsorption on silicon surfaces\cite{Cir83PRB}. In addition, 
it allows to describe the interactions between several different species within a uniform and coherent scheme. 
\textit{Ab initio} methods have the same advantage, but the systematic study of several manipulation conditions 
with a realistic system is still an unfeasible task. In our calculations, the silicon surface is represented by a four-layer 
cluster, each layer encompassing 4$\times$4 Si atoms (eight Si dimers). Hydrogen atoms are 
added to passivate the bottom and the sides of the slab, leaving a clean Si(001) surface on top (see Fig.~2). Careful 
checks have shown that this cluster size is large enough to avoid 
unwanted size effects. Relaxation of the bare Si(001) surface with the ASED method leads to a symmetric (2$\times$1) surface  
instead of the c(4$\times$2) reconstruction. However, the energy difference between these two structures is very small \cite{Ram95PRB}
and within the uncertainty of the ASED model. In addition, the height difference between the `up' and `down' atoms of the 
asymmetric dimer atoms is also small. Therefore, we expect correct tendencies for manipulation processes involving much larger energies and amplitudes.
The tip apex has been modeled using Si and Au clusters built along the (111) orientation, with 13~Si or 10~Au 
atoms for sharp tips (see Fig.~7). 
Each Si atom located in the base layer of the pyramidal tip was saturated by an hydrogen. All the EHMO parameters used in 
our calculations are reported in the table~\ref{EHMOparam}.

With the second model, we describe the long-range van der Waals forces between the tip and the surface. Although these interactions 
are the leading forces for the manipulation of physisorbed atoms\cite{Bou93PRB,Bou97PRB,Bou99PRB}, they 
are expected to have a negligible effect in the case  
of a chemisorbed atom. We can then safely neglect the long-range interactions involving the Ag atom, as 
well as the very small contributions arising from  
the tip cluster apex. However, it is essential to take into account the van der waals interaction between the macroscopic tip and 
the substrate in order to get a correct force distance relation\cite{Tan95EPL}. It is well known that 
quantum chemistry methods such as ASED or LDA-related techniques are unable to describe the dispersive van der Waals interactions. Here we have 
used an analytical Hamaker model \cite{Arg96JAP}, with the parametric relation (Eq~\ref{form_vdw}) 
derived from a tip built with an infinite cone \cite{notevdw} and 
a spherical cap, and an infinite surface. The geometry of the model is shown on the left panel of the Fig.~2. We used 
a spherical cap radius $\rho=50$~nm. The Hamaker constant $A$ for the Si-Si interaction is $1.865\times10^{-19}$~J \cite{Sen95CS}. In (\ref{form_vdw}), 
$h$ is actually the distance between the surface and the bottom of the macroscopic tip (without the tip apex cluster), i.e. 
$h$ is the sum of the distance $Z$ 
between the surface and the bottom of the tip apex cluster, determined from the center of the atoms, and 
of the height of the cluster (about 6.4~\AA\ for the Si(111) 
tip). With this parametrization, we obtained 
a maximum attractive van der Waals force of 6.42~nN between the Si tip and the Si surface. 

\begin{eqnarray}
F_z(h) & = & {A\rho^2(1-\sin\gamma)(\rho\sin\gamma-\rho -h\sin\gamma-h) \over 6h^2(\rho+h-\rho\sin\gamma)^2} \label{form_vdw}\\ 
\nonumber & & - {A\tan\gamma(h\sin\gamma+\rho\sin\gamma+\rho\cos(2\gamma)) \over 6\cos\gamma(\rho+h-\rho\sin\gamma)^2}
\end{eqnarray} 

In view of the very slow displacement of the tip apex usually achieved experimentally in manipulation as compared 
to the relaxation time of the adsorbate on a surface, a molecular mechanics approach was preferred to a molecular 
dynamics approach \cite{Tan97SS}. For each tip position (X$_{tip}$,Z$_{tip}$) considered, the Ag and slab atoms positions 
are optimized to get the system minimum energy. X$_{tip}$ is the tip apex coordinate along the Si dimer row axis, a zero value referring to 
the initial location of the silver adatom. In order to avoid time-consuming calculations, only the Ag atom and 
all the surface silicon atoms closer to Ag than a fixed distance R (see Fig.~2) were allowed to relax. We have checked the validity of 
this approximation by considering several R distances, from 5~\AA\ to 0~\AA\ (frozen surface). In most of the cases, we used a frozen surface 
because we got no qualitative changes and non significant quantitative variations. The tip atoms are not allowed to relax, what could be 
questionable for small tip surface distances. Consequently, our tip was always located high enough to ensure no significant interaction 
with the surface atoms. In this work, we have considered constant height and constant force mode for the manipulation. The simulation 
of the constant height mode is 
straightforward; for a given Z$_{tip}$, we performed calculations for a large range of X$_{tip}$ values ($\Delta X_{tip}=0.12$~\AA\ between two steps). 
The constant force mode is far more complicated and rises several issues which will be discussed in the last section.  

\section{Constant height manipulation}

In this section, we describe the results obtained in the constant height mode. In particular, we focused on the conditions 
required for adatom manipulation. Different kind of adatom movements have been observed, depending on the tip height, orientation, shape, or chemical
nature. However, we considered that a successful manipulation occurred only when 
the tip translation by a surface lattice parameter leads to an adatom movement in the same direction, and 
when this process can be successively repeated. 

We have been able to obtain four different manipulation mechanisms with our calculations. One is a vertical process where the adatom transfers 
from the surface to the tip. The three others are the usual lateral manipulation processes, sliding, pulling and pushing \cite{Mey97SS}.
The upper panel of the Figure~3 shows selected snapshots of the first case. When the tip is still far away from the adatom 
(on the left of the Fig.~3), one can see that the height difference remains constant and that the X-position 
difference decreases continuously, i.e. the adatom stays in the cave site, unperturbed by the tip approach. The energy does not vary, 
meaning that the tip is high enough such that the tip-surface interaction is negligible. At a specific tip position (position (a) in Fig.~3), 
which depends on the tip height, the configuration becomes energetically unstable, and the adatom ``jumps'' under the tip apex (position (b) 
in Fig.~3). Obviously, we do not have a lateral manipulation since the adatom does not remain on the surface. 
The mechanism is similar to the one depicted for the transfer of one atom between two flat electrodes \cite{Bul96PRB}. There are two 
adsorption minima, one in the cave site and the other under the tip apex, separated by an energy barrier. Because our calculations are done with T$=0$~K, 
the transition occurs only when the energy barrier vanishes. The transition is clearly visible in the curves, the X-position difference 
being now almost zero, and the height difference being reduced 
by 0.9~\AA. The energy difference between the two minima when the tip is far away from the surface, is about 260~meV in favor of adsorption under the tip.
It is energetically favored because the adatom is strongly bonded with the dangling orbital on the apex atom of the Si(111) tip. 
After the transition, the Ag adatom remains attached to the tip apex (positions (b) and (c) in Fig.~3). Weak undulations of 
the X-position difference and of the energy prove that the Ag atom 
still ``feels'' the surface. The variation of height and energy are perfectly sinusoidal, which Bouju et al have shown to be impossible 
in the case of a sliding mode, i.e. with the adatom remaining on the surface \cite{Bou99PRB}. The adatom is now clearly bonded only to the tip. 
Therefore, if the tip is raised, the Ag atom is carried away. 
However, subsequent deposition may follow if the adatom is brought by the tip in the vicinity of a surface 
site with a lower adsorption energy, near a step or a kink for instance. 
Such investigations are beyond the scope of this study.

Lateral manipulation occurred only within three distinct processes \cite{Mey97SS}. In the sliding mode, the adatom is trapped by the tip apex and 
follows the tip displacements \cite{Eig90NAT,Mey97SS,Zep92ULT,Bar97PRL}. It can be obtained if the tip is closer to the surface (Fig~4). During 
the tip approach, the adatom jumps again under the apex, and is subsequently trailed by the tip. Unlike the surface-to-tip transfer case, the 
adatom is still chemisorbed on the surface. Whether or not the tip is more attractive than the surface is no more important in this tip height 
range, since there is now only one minimum energy position between the tip and the surface. 
The adatom is pushed back both by the tip 
and the surface. The only condition for the tip trailing the adatom is that the potential trap is deep enough to overcome the diffusion barrier on
the surface. The X$_{Ag}$-X$_{tip}$ variation represented in the Figure~4 shows the expected characteristics of a stick-slip behavior \cite{Bou99PRB}, 
also very well known in the field of friction \cite{Mey92SPR,Mar93NATO,Gya95EPL,Tom96SPR,Car97CR,Hol97SS,Shi98NANO}. 

Another manipulation process is the pushing mode where the adatom is repelled by the tip displacement. In the Figure~5, an example of pushing 
is represented for another configuration including a rotated Si(111) tip and the Ag atom initially located in the bond site. Now, the X-position 
difference ranges from 3 to 4~\AA, indicating that the Ag atom is always in front of the moving tip. At the beginning, the distance between the tip 
and the adatom decreases, since this one prefers to remain in the initial bond site (position (a) in Fig.~5). It then jumps in the next bond site 
(position (b) in Fig.~5), and the process is repeated (position (c) in Fig.~5). The movement of the silver adatom does not appear in the 
energy curve (Fig.~5). In fact, the energy difference between the two configurations before and after the transition is very small. 
Instead this is the variation of the tip-surface interaction energy which dominates the curve. The asymmetry of the curve can be explained by the 
asymmetry of the tip.

In the last process, the manipulated object is pulled by the scanning probe tip (Fig.~6). During the first tip steps, we get a behavior similar that for 
the sliding mode. The tip is approaching the silver atom in the cave site, and at some point, this one jumps under the tip (position (a) in Fig.~6). 
However, instead to be right under the tip apex, it is now located between the cave site and the tip apex. The tip is too low to allow the adatom to remain 
permanently under the tip apex. Then, the energy continues to decrease, with a minimum around X$_{tip}=0$ where the tip apex and the adatom are right 
above the case site. It means that the Ag atom is attracted by both the tip and the surface. Further tip steps lead to an increase in energy whereas 
the adatom prefers to remain in the vicinity of the cave site. At some point, this configuration becomes unstable, and the adatom jumps toward 
the rear of the tip (position (b) in Fig.~6). We will show later how this behavior can be correlated with the tip geometry. The adatom is now pulled by the 
tip during the scan. One can see the non-sinusoidal undulations in the X-position and height differences. The silver adatom feels both the attraction of the 
tip and of the supporting Si(001) surface. The process includes two distinct steps. In the first one, the adatom goes up and tries to stay close to 
the stable cave site (position (c) in Fig.~6). When the tip attraction exceeds the surface attraction, the adatom quickly comes again close to the tip rear. 
This mode is somewhat peculiar, because the tip has to pass over the adatom in order to trail him. However, the same result could be achieved 
if the tip is first high above the surface, then lowered in front of the adatom, before the beginning of the scan. 

In the Figure~7, we have summarized the results obtained for several cases of constant height scans, with a tip height ranging from 
0.5~\AA\ to 6~\AA, with successive increments of 0.1~\AA (Si tips) or 0.5~\AA (Au tips).
With the Ag atom initially in the cave site, we have 
first considered a Si(111) tip (A), then the same Si tip, but rotated by 120$^\circ$ to have a leading edge (B), an Au(111) tip (C), and finally a 
truncated Au(111) tip without Au atom at the apex (D). With the Ag atom initially in the bond site, the rotated Si(111) tip was first used (E), then 
an Au(111) tip (F), and the truncated Au(111) tip (G). Constant height scans 

The different areas represent Z$_{tip}$ ranges for five different adatom behaviors.
Besides the four manipulation processes depicted above, two other cases are emphasized.
In the white areas in the Figure~7, the tip passes over the adatom which remains in its initial location, because either the tip 
is too high or the tip attraction on the adatom is not strong enough to overcome the diffusion barrier. Instead, checkerboard areas gather 
different unsuccessful behaviors, where the adatom typically avoids the tip by jumping in nearest lateral adsorption sites. Our calculations 
have been done at T$=0$~K, and one can expect such site hopping's in a larger tip height range in experiments, because of thermal enhancement. This 
argument emphasizes the need to perform atomic manipulation experiments in a very low temperature environment. 

A preliminary study had shown that the manipulation of the Ag adatom on the Si(001) surface was feasible under specific conditions. Indeed, it has been 
calculated that with a gold tip, pushing occurs within a tip height range between 0.2 to 0.5~\AA\ \cite{Piz97SSL}. This range is may be too small 
to be certain of the transferability of this result in the experiments, where conditions could be different. 
However, the overview of our results depicted in the Figure~7 clearly shows that the tip height ranges of successful simulations are typically between 1.5 and 
3~\AA\ for the Si tip. This is large enough to ensure that our results are not strongly dependent on the physical approximations we have made. In addition, the
controlled manipulation is predicted for different kind of tips and adsorption sites. This is a crucial requirement since despite recent successes in in-situ 
structural characterization of the very end of the tip \cite{Cro98PRL}, the geometry of the tip apex is usually not known. Therefore, our results show 
that with scanning probe techniques, one should be able to manipulate one Ag atom on the Si(001) surface.

Several clues about the manipulation on surfaces like Si(001) can be drawn from the calculations. Here, we found that a sliding process is possible when the 
tip is made in silicon but not in gold. This result can be explained by the large diffusive energy barrier to overcome on the Si(001) surface. The potential 
well generated by the Si tip is deep enough to drag the adatom on the surface which can not be done with an Au tip. In this last case, only the core-core 
repulsive pushing mode is possible. Another necessary condition to perform manipulation is that the adatom stay in line with in the tip direction. On a surface 
strongly anisotropic, vicinal with regular steps for instance, this condition can be fulfilled whatever 
the tip. However, for the Si(001) surface, the adatom must also be guided
by the tip. Indeed, the Si tip, with the directional bondings originating from the `s-p' hybridization, prevents in most cases the adatom to escape aside. On 
the contrary, the Au tip has a more isotropic adsorption energy surface due to its predominant `s' bonding. It explains why its range of successful pushing is 
so narrow, the adatom jumping sideways or staying low below the tip.   

In the Figure~8, we have represented the position and adsorption energy of the Ag adatom on the isolated Si tip, along the scan direction. We noticed that during 
the manipulation processes, the Ag atom was always located in or very close to a position corresponding to an energy minimum of the energy curve. The minimum 
labeled A corresponds to the surface-to-tip transfer or sliding mode (Ag under the tip), B to the pulling mode (Ag behind the tip) and C to the pushing mode (Ag 
in front of the tip). Our results also suggest that the ranges associated with different processes in the Figure~7 depend on the characteristics of these minima.
The column (A) of the Figure~7 shows that sliding and pulling occur for large ranges. Accordingly A and B are deep minima, where the adatom can be accommodated 
over a large variation of tip height. Conversely, C is a very small minimum, almost an instable point, and it corresponds to the narrow range of tip height of 
the pushing mode. Further support for this relationship come by considering the column (B) of the Figure~7. Because the Si(111) tip is invariant under the C$_3$
symmetry operation, the Figure~8 also shows the 120$^\circ$ rotated Si(111) tip if the scan is towards decreasing X. Accordingly, B is now the minimum where the Ag
atom is adsorbed during the pushing mode and C during the pulling mode. One clearly see that because B is a deep minimum, the pushing mode is now possible 
over a large range. Our suggestion is also supported by the result shown in the column (E) of the Figure~7, where pushing and sliding mode occur over large similar
tip height range. In this last case, the Ag atom is now located in the bond site, which tends to indicate that this criterion may be general whatever the surface,
the tip or the kind of adatom. However, one must be aware that these observations remain essentially qualitative and limited to a set of calculations. The success
or the failure of the manipulation is decide by non-trivial interactions between the surface, the adatom and the tip. Our results seem to show that the tip is the
essential part of the trio, but more investigations are needed to confirm this phenomenological criterion.

\section{Forces signatures}

Experiments done in the group of Pr. K.-H.~Rieder have highlighted that, during constant current STM manipulation of adatoms, recognizable patterns in the response 
of the tip height signal can be associated with each manipulation process \cite{Bar97PRL,Bar97CPL}. The variations pertaining to each process have been recently
reproduced using a numerical simulation of an STM tip and its associated feedback loop device \cite{Bou99PRB,Bou01PRB}. During the manipulation, the object switches between 
positions alternatively close or far to the immediate vicinity of the tip apex, what is responsible for the large variations of the extremely sensitive tunneling
current. The perturbations caused by the object will not only modify the tunneling current but also the forces acting on the probing tip. We therefore expect to
find similar patterns in the force monitored during constant height scans. 

We have calculated the $x$ and $z$ components of the force \textbf{F} acting on the scanning probe tip, using the computed energy values during a large number of
constant height scans. Once the energy is known on each node of a fine \{X$_{tip}$,Z$_{tip}$\} mesh, the forces are derived from the usual formula. For the derivation 
of F$_z$, care has been taken to use tip height values close to each other and corresponding to the 
same manipulation mode. The results for one scan at Z$_{tip}=3.6$~\AA\ are
represented in the Figure~9. When the tip is far from the adatom (left part of the Figure~9), \textbf{F} and the tip-surface 
force \textbf{F}$^0$ are identical and show negligible
variations since the tip is too high to feel the corrugation of the surface. F$_z$ and F$_z^0$ are not zero because of the long-range attractive van der Waals contributions.
The force on the tip deviates from zero as soon as the distance with the adatom is slightly smaller than a lattice parameter (3.84~\AA). F$_z$ decreases whereas 
F$_x$ increases, the tip being attracted by the adatom. When X$_{tip}=-1.5$~\AA, the curve F$_z$(X) and F$_x$(X) show discontinuities related to the transfer of 
the adatom under the tip. These divergences originate from the numerical calculations of the energy derivatives. Once the adatom 
located under the tip, we observe oscillations of both F$_z$ and F$_x$, with the surface lattice periodicity. Actually, the tip 
is now probing the silicon surface, since the tip is enlarged by the adatom
at its apex and is then closer to the surface. The transfer of one adatom under the tip can then be recognized from the increased sensitivity of lateral and
vertical forces. An equivalent situation is the sudden increase of contrast obtained with a STM when an adatom initially on the surface transfers under the tip. 

The same analysis has been performed for all the configurations considered in the previous part. To get insights about the force variations 
involved during the manipulation, it is required that F has a
much larger amplitude than F$^0$. The Figure~10 shows some of the most significant force signatures we have obtained. The upper panel A represents the vertical and
lateral forces signatures after the adatom being transferred under the tip. F$_z$ minima occur when the tip carrying the adatom is exactly above the cave sites. 
Here F$_z$ is lower than F$_z^0$, which indicates that the force between the tip and the adatom is attractive. F$_x$ is small and oscillate around zero, the adatom being
successively attracted by the cave sites in line on the surface. Neither F$_z$ nor F$_x$ show the typical sawtooth behavior associated with the lateral 
manipulation. The adatom must stay on the surface in order to introduce an asymmetry in the forces variation. The surface-to-tip transfer can then only be 
recognized by the increased sensitivity of the tip. The central panel B shows the forces variations calculated during a sliding process as depicted in the 
Figure~4. First, one can see that F$_z$ is greater than F$_z^0$, so the vertical force between the adatom and the tip is repulsive. The most apparent feature of the
forces variation is the apparition of asymmetric sawtooth patterns for F$_x$(X), and also for F$_z$(X), though more weakly. The analysis of the curves in relation with the
adatom positions show the same kind of stick-slip behavior obtained in STM experiments \cite{Bar97PRL,Bar97CPL} and also well known in the theory of 
friction \cite{Mey92SPR,Mar93NATO,Gya95EPL,Tom96SPR,Car97CR,Hol97SS,Shi98NANO}. Starting 
with both adatom and tip at the cave site position (F$_x(a)=0$ and F$_z(a)$ close to a minimum), 
F$_x$ becomes negative and decreases slowly as the tip moves away, the tip being ``retained'' by the adatom which tries to remain in the cave site. When the tip is
approximately 1.1~\AA\ ahead, the adatom position is instable and it quickly reaches a new position between the tip and the following cave site. F$_x$ suddenly
becomes positive whereas the peak in F$_z$ corresponds to the tip and adatom between two cave sites, in a highly repulsive configuration. F$_z$ and F$_x$ then
slowly decrease until both the tip and the adatom are again in a cave site. Similar analysis can be done for a pushing process (panel C of the Figure~10). Here, although
F$_x$ shows a sharper but similar variation than in the B panel, the asymmetry in the oscillation of F$_z$ is now reversed. The slow increase of F$_z$ is associated with
the approach of the tip, the contribution of the adatom to F$_z$ being more and more repulsive, and the fast decrease of the force is now caused by the escaping
jump of the adatom in direction of the next bond site. 

Constant current mode STM studies have shown that it was possible to extract the manipulation modes from the tip height variations \cite{Bar97PRL,Bar97CPL}. The
Figure~10 suggests that the analysis of forces variations could also provide similar and meaningful insights. Tunneling current and interactions are completely
different quantities, but it is possible to compare their variations as a function of the tip-surface distance. The tunneling current is a positive
scalar quantity, and it approximately changes by one order of magnitude 
for one angstr\"{o}m, whatever the nature of the tunneling junction. The distance variation is far
more complicated for interaction forces, which can be repulsive or attractive, the latter with either positive or negative gradient. The gradient depends on the
nature of the materials \cite{Cro98PRL,Ros81PRL}, and the shape of the tip is also known to have a non-negligible influence \cite{Cir90PRB,Nes93SSL,Per97PRL}. In
the case of a blunt W tip in interaction with an Au(111) surface, the force gradient has been measured \cite{Cro98PRL} between $-4$~nN.\AA$^{-1}$ and 
$+1$~nN.\AA$^{-1}$. We therefore expect that the force signatures might show approximate sawtooth shape, unlike the height variations in the STM measurements. 

Using our results and deduction, we gathered in the table~\ref{signs} the force signatures, just as the tip height signatures reported in STM constant
current experiments, for
all the possible cases. We distinguished two cases whether the force gradient is positive or negative. It is to be noted that two cases must also be considered for
the tip height variation in the STM experiments. In fact, it is 
well known that most of the adsorbates will appear as a protrusion in the constant current image, effectively increasing the
tunneling current, but others will act as a kind of screen for the tunneling and then appear as a depression in the image \cite{notetunn}. The variation shape are
then inversed, as been shown in STM experiments \cite{Bar97PRL}. In our calculations, we were unable to obtain the forces variations for both the sliding/pulling
and pushing processes in the positive gradient regime. The adsorption energy minimum being lower on the Si tip, a vertical transfer happened in the relevant tip
height range. The strong attraction of our tip, needed to overcome the large diffusion barrier of the surface, also explains why the calculated forces 
signatures are not as sharp as in the table~\ref{signs}. During the manipulation, the adatom remained most of the time in the proximity of the apex, and the lateral jumps to
the next site were small. Larger jumps would greatly enhance the sharpness of the signature patterns. 

The forces signatures compiled in table~\ref{signs} could first help to determine if manipulation occurred. Indeed, a sudden increase of the asymmetry of F$_x$ variation
means that the tip push or drag an object on the surface. Also, the increase of the amplitude of the F$_z$ variation may be associated with a transfer of the
object under the tip. Manipulation processes could also be recognized from the analysis of the F$_z$ sawtooth variations. Once the relation between the force and
the tip-surface distance is determined, i.e. we know if the gradient is positive or negative, it should be relatively easy to deduce if sliding/pulling or pushing
had happened during the scan.

\section{Constant force manipulation}

Theoretical studies of the manipulation of one atom by a scanning probe tip usually focused on the constant height mode. Its simulation is easier than for 
modes based on the control of one interaction quantity, such as constant current in STM or constant force in AFM. Indeed, in constant height 
mode, there is no need to take into account the effect of a feedback loop on the scanning tip. To our knowledge, the first attempt to 
simulate other modes has been achieved by Bouju et al \cite{Bou99PRB}. With their \textit{virtual} scanning tunneling microscope, they 
were able to reproduce the sawtooth variation of the STM tip height during the constant current manipulation of a Xe atom on a Cu(110) surface. 
However, as far as we know, other modes, such as AFM constant force, have not been investigated yet in the purpose of manipulation. From the experimental 
point of view, no reports of manipulation in constant force mode have been made. The occurrence of the so-called snap-to-contact due to mechanical 
instabilities when the force gradient $\partial$F$_z/\partial z$ exceeds the cantilever stiffness for small tip surface separation may be 
one of the inherent difficulties \cite{notefeed}. Therefore, numerical simulations are an 
invaluable tool to investigate the manipulation mechanisms in this mode. 

The straightforward way for simulating constant force scans consists in implementing an extra algorithm in the computational code, which mimic 
the feedback loop. First, we 
have tried this approach. We made the assumption that the characteristic time of the adatom relaxation is much smaller than 
the feedback loop response, i.e. the time required to adjust the height of the tip. Practically, the adatom position is then fully relaxed for each 
tip height, which corresponds to the adatom following adiabatically the tip vertical movement. A low tip speed adjustment is needed in order to avoid 
large oscillations of the tip, which may lead to tip crashing or lateral adatom jumps. Although less elaborate, this scheme is equivalent to the 
technique used for the \textit{virtual} scanning tunneling microscope \cite{Bou99PRB}. However, this automatic technique shows several 
drawbacks for constant force mode. 
In the useful range, the variation of the force between the tip and the surface as a function of the tip height is not monotonic, unlike 
the variation of the tunneling current with the distance. This problem can be tackled by a careful optimization of the speed of the tip height adjustment. 
Hence, we were able to obtain constant force scans of the bare surface using this technique. However, additional difficulties appear when the scanning 
tip interacts with the supported adatom. For example, a jump of the adatom directly under the tip may drastically change the interaction 
between the tip and the surface. In the tip height range useful for the manipulation, an attractive vertical force 
will be repulsive after the jump, which will result in instabilities in the feedback loop. 

Therefore, in this work, we favored another approach, with which these instabilities are clearly identified. Practically, we used the 
vertical force between the tip and the surface calculated for each {X$_{tip}$,Z$_{tip}$} on a fine grid. The sequence of the tip heights Z$_{tip}$ 
during a scan is then extracted for successive X$_{tip}$ from this set by assuming a constant force. When an instability happens, for example 
because of an adatom jump under the 
tip apex, we then consider a new set of positions and forces associated with the new tip configuration, i.e. with the adatom located under 
the tip. The advantage of this method is the possibility to get all the solutions for a given force. In particular, we can analyze precisely the 
behavior of the system near a feedback instability. 

The Figure~11 shows tip height variations calculated for several force values, in the case of a Si(111) tip and the Ag adatom in the cave site. 
Far from the neighborhood of the adatom, we have obtained height variations with the periodicity of the Si(001) substrate, with asymmetric shapes reflecting 
the non-symmetric configuration of the tip. The minimum heights occurred close to the cave sites locations. For F$_z=-1.6$~nN, the tip still did not ``feel'' the  
adatom until X$_{tip}$ is approximately lower than $-a$. After this position, the tip rose because of the increasing interaction with the adatom. The tip 
then passed over the adatom with Z$_{tip}=4$~\AA, high enough to prevent a surface-to-tip transfer. The higher bump at X$_{tip}=0$ is the image of the Ag 
atom in the constant force mode. However, for F$_z=-1.8$~nN or F$_z=-2.0$~nN, the tip is closer to the
surface during the scan, and the adatom jumped under the tip apex when X$_{tip}\simeq -a/2$. In the normal conditions, the slow motion of the tip holder 
causes gradual changes. But here, the adatom jump completely changes the value of F$_z$, and mechanical instabilities are expected to follow.

Now we focus on the effect of the surface-to-tip jump of the adatom during the scan at F$_z=-1.8$~nN. Owing to the 
much longer response times of the cantilever and the feedback, the 
final outcome can be determined from the static F$_z(z)$ at the location of the adatom jump. 
We reported in the Figure~12 the possible tip trajectories. The left graph shows the tip height variation before the jump, reported from the Figure~11. At the jump 
(X(tip)$=-1.72$~\AA), the adatom is transferred under the tip. The right graph of the Figure~12 represents how the tip height varies as a function of the 
force at the transition point. The curve~A corresponds to the configuration with the adatom on the cave site, whereas for B and C, the adatom is now located at the tip
apex. The intersection of the curves with the dashed line gives the tip height for each configuration. After the jump, there are two solutions, one for the 
branch B (negative force gradient) and the other for the branch C (positive force gradient). We have to examine the two possible transitions 
A$\to$B and A$\to$C. The corresponding continuations of the scan are shown in the left graph of the Figure~11. First, we considered the transition A$\to$B.  
At B, the force gradient is negative, unlike the configuration A. Because the feedbacks are usually assumed by design to keep the sign of the force 
gradient, a slow instability of the tip holder is expected, followed by a crash on the surface. The other possible transition, A$\to$C, does not suffer 
the gradient restriction. At A, before the adatom jump, the tip-cantilever system is in equilibrium, the attraction by the surface being balanced 
by the return force of the cantilever. Right after the adatom transfer, 
the tip-surface force becomes suddenly repulsive and add up with the return force to rise the tip. The tip will oscillate around the new 
equilibrium position with a maximum amplitude equal to the difference between the initial and final positions. Here, the height difference between A and C is about 
1~\AA. If we assume that the adatom remains in its location at the tip apex during the oscillations and the that the feedback speed is much greater than the scan 
speed, these oscillations will be progressively dampened and the slow compensation by the feedback loop will bring the tip to the new equilibrium position. 
The scanning will then continue with the height variations shown on the branch C in the left graph of the Figure~12. 

For the system we have considered, in the range of attractive force with a positive gradient, we observed that either the tip images the 
adatom or a surface-to-tip transfer happened during the scan over the adatom. Other possibilities, 
encompassing attractive forces with negative gradient, repulsive forces, or different tip shapes and orientations, have not been fully 
investigated, an exhaustive study being out of the scope of the present paper and could be the matter of a future publication. However, we emphasize that our analysis 
remains valid for the general case since we have always observed a jump of the adatom during the approach of the tip.

\section{Conclusion}

We have investigated the manipulation by a scanning probe tip of one silver atom deposited on a Si(001) surface using a molecular mechanics approach. The system 
was described by a cluster model, and we have used the combination of a quantum chemistry model for the short-range interactions and an analytical Hamaker 
model for the long-range van der 
Waals contributions. We have identified four different possible manipulation processes, i.e. the surface-to-tip transfer, the sliding, pulling and pushing modes, 
when using a Si(111) tip, whereas only pushing is achieved in the case of an Au tip. The structure of the tip apex has been shown to be the predominant factor for 
determining the ordering and tip height ranges associated with each manipulation process. The second part of the paper was devoted to an original study of the forces 
variations during the manipulation. We have determined that each process is associated with characteristic variations, very similar to the height 
signatures observed in constant current STM experiments. It is suggested that the kind of process involved in the manipulation could be 
inferred from the monitoring of the forces. 
Finally, we have presented the first numerical simulations of the AFM constant force mode. From a preliminary analyze, the different possible 
behaviors of the tip during a surface-to-tip transfer are discussed, taking into account the feedback and cantilever responses.

\begin{acknowledgments}

We are grateful to C.~Joachim, E.~Meyer, and H.~Hug for fruitful discussions, and to X.~Bouju for critical reading of the manuscript. 
We acknowledge the Swiss National Foundation for financial and computational support under the 
program NFP36 ``Nanosciences''. Part of the computations 
have been done at the Centre d'\'{E}laboration des Mat\'{e}riaux et d'\'{E}tudes Structurales (Toulouse, France). 
\end{acknowledgments}


\begin{thebibliography}{88}
\bibitem[{not({\natexlab{a}})}]{notevdw}
\bibinfo{note}{The cone angle is $\gamma=0.5137$~rad. With this value, the cone
  shape is equivalent to a pyramidal shape with an aspect ratio 1:1 (see ref
  \cite{Arg96JAP}).}
\bibitem[{not({\natexlab{d}})}]{noteperp}
\bibinfo{note}{During the manipulation, the adatom remained almost in line with
  the tip, and then we have not represented the variation of the adatom
  position in the direction perpendicular to the tip movement.}
\end{thebibliography}

\newpage 

\section*{Table captions}

\begin{table}[h]
\caption{EHMO parameters. $\zeta_{s,p,d}$ are the Slater orbital exponents, IP$_{s,p,d}$ the ionization potentials and 
C$_{1,2}$ the linear coefficients for double-$\zeta$ functions.}\label{EHMOparam}
\end{table}

\begin{table}[h]
\caption{Tip height (constant current scans) and forces signatures (constant height scans) during manipulation processes (sliding/pulling and pushing) with a 
scanning probe tip. We distinguished two cases whether the manipulated object is seen as a depression with STM and the force F$_z$ between the tip and the 
object varies with a positive gradient (case~$\bigvee$), or the object is seen as a protrusion with STM and the force F$_z$ varies with a negative gradient
(case~$\bigwedge$).}\label{signs}
\end{table}

\clearpage

\ \vspace{4cm}

\begin{table}[h]
\begin{tabular}{cccccccccc}
& $\zeta_s$ & IP$_s$ & $\zeta_p$ & IP$_p$ & $\zeta1_d$ & IP$_d$ & C$_1$ & $\zeta2_d$ & C$_2$ \\ \hline
H & 1.1600 & -13.600 & & & & & & & \\
Si & 1.6998 & -13.460 & 1.4855 & -8.151 & & & & & \\
Ag & 1.9060 & -7.580 & 1.6200 & -3.920 & 4.9890 & -10.500 & 0.5576 & 2.5840 & 0.5536  \\
Au & 2.6020 & -10.920 & 2.5840 & -5.550 & 6.1630 & -15.070 & 0.6851 & 2.7940 & 0.5696 \\
\end{tabular}
\end{table}

\vfill\centerline{\bf Table 1}

\clearpage

\ \vspace{4cm}

\begin{table}[h]
\begin{tabular}{ccccc}
& & Tip height & F$_z$ & F$_x$ \\ \hline
$\bigvee$ & sliding/pulling & 
\unitlength=0.5mm
\begin{picture}(45,15)(0,0)
\put(5,1){\line(1,1){10}}
\put(15,11){\line(0,-1){10}}
\put(15,1){\line(1,1){10}}
\put(25,11){\line(0,-1){10}}
\put(25,1){\line(1,1){10}}
\put(35,11){\line(0,-1){10}}
\end{picture}
&
\unitlength=0.5mm
\begin{picture}(45,15)(0,0)
\put(5,1){\line(1,1){10}}
\put(15,11){\line(0,-1){10}}
\put(15,1){\line(1,1){10}}
\put(25,11){\line(0,-1){10}}
\put(25,1){\line(1,1){10}}
\put(35,11){\line(0,-1){10}}
\end{picture}
&
\unitlength=0.5mm
\begin{picture}(45,15)(0,0)
\put(5,11){\line(1,-1){10}}
\put(15,1){\line(0,1){10}}
\put(15,11){\line(1,-1){10}}
\put(25,1){\line(0,1){10}}
\put(25,11){\line(1,-1){10}}
\put(35,1){\line(0,1){10}}
\end{picture}
\\
& pushing & 
\unitlength=0.5mm
\begin{picture}(45,15)(0,0)
\put(5,11){\line(1,-1){10}}
\put(15,1){\line(0,1){10}}
\put(15,11){\line(1,-1){10}}
\put(25,1){\line(0,1){10}}
\put(25,11){\line(1,-1){10}}
\put(35,1){\line(0,1){10}}
\end{picture}
&
\unitlength=0.5mm
\begin{picture}(45,15)(0,0)
\put(5,11){\line(1,-1){10}}
\put(15,1){\line(0,1){10}}
\put(15,11){\line(1,-1){10}}
\put(25,1){\line(0,1){10}}
\put(25,11){\line(1,-1){10}}
\put(35,1){\line(0,1){10}}
\end{picture}
&
\unitlength=0.5mm
\begin{picture}(45,15)(0,0)
\put(5,11){\line(1,-1){10}}
\put(15,1){\line(0,1){10}}
\put(15,11){\line(1,-1){10}}
\put(25,1){\line(0,1){10}}
\put(25,11){\line(1,-1){10}}
\put(35,1){\line(0,1){10}}
\end{picture}
\\ \hline
$\bigwedge$ & sliding/pulling & 
\unitlength=0.5mm
\begin{picture}(45,15)(0,0)
\put(5,11){\line(1,-1){10}}
\put(15,1){\line(0,1){10}}
\put(15,11){\line(1,-1){10}}
\put(25,1){\line(0,1){10}}
\put(25,11){\line(1,-1){10}}
\put(35,1){\line(0,1){10}}
\end{picture}
&
\unitlength=0.5mm
\begin{picture}(45,15)(0,0)
\put(5,11){\line(1,-1){10}}
\put(15,1){\line(0,1){10}}
\put(15,11){\line(1,-1){10}}
\put(25,1){\line(0,1){10}}
\put(25,11){\line(1,-1){10}}
\put(35,1){\line(0,1){10}}
\end{picture}
&
\unitlength=0.5mm
\begin{picture}(45,15)(0,0)
\put(5,11){\line(1,-1){10}}
\put(15,1){\line(0,1){10}}
\put(15,11){\line(1,-1){10}}
\put(25,1){\line(0,1){10}}
\put(25,11){\line(1,-1){10}}
\put(35,1){\line(0,1){10}}
\end{picture}
\\
& pushing & 
\unitlength=0.5mm
\begin{picture}(45,15)(0,0)
\put(5,1){\line(1,1){10}}
\put(15,11){\line(0,-1){10}}
\put(15,1){\line(1,1){10}}
\put(25,11){\line(0,-1){10}}
\put(25,1){\line(1,1){10}}
\put(35,11){\line(0,-1){10}}
\end{picture}
&
\unitlength=0.5mm
\begin{picture}(45,15)(0,0)
\put(5,1){\line(1,1){10}}
\put(15,11){\line(0,-1){10}}
\put(15,1){\line(1,1){10}}
\put(25,11){\line(0,-1){10}}
\put(25,1){\line(1,1){10}}
\put(35,11){\line(0,-1){10}}
\end{picture}
&
\unitlength=0.5mm
\begin{picture}(45,15)(0,0)
\put(5,11){\line(1,-1){10}}
\put(15,1){\line(0,1){10}}
\put(15,11){\line(1,-1){10}}
\put(25,1){\line(0,1){10}}
\put(25,11){\line(1,-1){10}}
\put(35,1){\line(0,1){10}}
\end{picture}
\\
\end{tabular}
\end{table}

\vfill\centerline{\bf Table 2}

\clearpage

\section*{Figure captions}

\noindent Figure~1: Adsorption energy map of one Ag deposited on Si(001)-2$\times$1.
The black spheres represent the Si surface and subsurface atoms. The thick arrow shows the tip movement 
direction.
\vspace{1cm}

\noindent Figure~2: Models used in the calculations. The left panel represent the geometrical shapes (light grey) from which 
long-range van der Waals forces are estimated. The small dark grey region shows the very end of the scanning tip, the adatom and the 
surface beneath which are treated by an extended H\"{u}ckel calculation to account for the chemical bondings. 
A ball-and-stick model of this region shows the clusters used to represent the surface and the Si(111) tip (right panel).
\vspace{1cm}

\noindent Figure~3: Example of a surface-to-tip transfer (Z$_{tip}=3.6$~\AA). Upper panel: snapshots for three 
different tip positions. The dark (light) grey spheres show the Si (Ag) atoms. 
Lower panel: variation of the X-position difference between the Ag adatom and the tip apex (empty squares), of the height difference  
between the adatom and the tip apex (filled circles), and of the short-range 
contributions to the total energy (thick line) \cite{noteperp}. The zero of 
energy corresponds to the relaxed 
system with the tip far away from the surface. Only 1 in 2 symbols (squares and circles) have 
been plotted for clarity. The vertical dotted lines mark the tip positions corresponding to the snapshots in the upper panel. 

\vspace{1cm}

\noindent Figure~4: Example of a sliding process (Z$_{tip}=2.7$~\AA). See the caption of the Figure~3 for explanations.

\vspace{1cm}

\noindent Figure~5: Example of a pushing process (Z$_{tip}=3.3$~\AA). Note that the Si(111) tip is here rotated and that the adatom is initially located in 
the bond site. See the caption of the Figure~3 for further explanations.

\vspace{1cm}

\noindent Figure~6: Example of a pulling process (Z$_{tip}=2.4$~\AA). See the caption of the Figure~3 for explanations.

\vspace{1cm}

\noindent Figure~7: Results of the scans for all the considered cases A-G. For A-D (E-G), the Ag adatom is initially located in the cave (bond) site. For each 
case A-G, the column represents the Z$_{tip}$ range corresponding to different adatom behaviors among five. The black color is for a successful pushing 
mode, the grey one for a surface-to-tip transfer or sliding mode, whereas shaded zones correspond to 
pulling processes. White and checkerboard areas shows unsuccessful 
manipulations. In the first case (white), the adatom stays in its initial location and in the second case (check board), the adatoms did not remain in line 
with the tip after the scan. In top (A-D) or bottom (E-G) of each column, we have represented a side (with the arrow) and a top views of the tip used in 
each case. White (grey) spheres represent Si (Au) tip atoms. 

\vspace{1cm}

\noindent Figure~8: (X,Z) position and energy of the Ag atom on the isolated Si(111) tip, the atom being at the center of the tip along the Y-axis (perpendicular to the 
scanning direction). The full circles show the tip atoms positions (tip apex at the origin) and the letters refer to different minima. The minimum A is the energy
reference. Note that if the tip 
scans to the decreasing X direction and not to the increasing X, it corresponds to the rotated Si(111) tip.  

\vspace{1cm}

\noindent Figure~9: Variation of the vertical (F$_z$) and lateral (F$_x$) forces acting on the tip as a function of the tip position along the dimer row direction for
Z$_{tip}=3.6$~\AA. The thick full lines show the forces \textbf{F}$_{z,x}$ with the adatom initially located at X$=0$, whereas 
the thick dashed lines represent the forces \textbf{F}$^0_{z,x}$ 
without the adatom. Also shown with thin lines are the differences (\textbf{F-F}$^0)_{z,x}$. $a$ is the lattice constant of the Si(001) surface ($a=3.84$~\AA).  

\vspace{1cm}

\noindent Figure~10: Variation of the vertical (F$_z$) and lateral (F$_x$) forces acting on the tip as a function of the tip position along the dimer row direction for three
manipulation processes. The full lines show the forces during the manipulation (\textbf{F}) whereas the dashed lines represent the forces without the 
adatom (\textbf{F}$^0$). The
first case A show a surface-to-tip transfer with an attractive vertical force between the adatom and the tip (Z$_{tip}=3.6$~\AA). The case B is an example of a
sliding process with a repulsive vertical force between the adatom and the tip (Z$_{tip}=2.8$~\AA). Finally C show a pushing process (Z$_{tip}=2.8$~\AA, rotated
Si(111) tip and adatom in site bond). 

\vspace{1cm}

\noindent Figure~11: Constant force scans for F$_z=-1.6$~nN, $-1.8$~nN and $-2$~nN (positive force gradient) with a Si(111) tip and the Ag atom in the cave 
site. The crosses mark the tip position at which the adatom jumps under the tip. Note that the adatom is initially located at X$_{tip}=0$. 

\vspace{1cm}

\noindent Figure~12: Left panel: Tip height variations as a function of the tip position for a constant force F$_z=-1.8$~nN scan, before and after the 
transition (marked by the dashed curve at X(tip)$=-1.72$~\AA). Right panel: Tip height variations as a 
function of the vertical force F$_z$ for the tip position X(tip)$=-1.72$~\AA. 
For the curves labeled by A, the adatom remains   
in its initial adsorption site (X(tip)$=0$~\AA), whereas it is located under the tip for B (negative force gradient, 
$\partial$F$_z/\partial z<0$) and C (positive force gradient, $\partial$F$_z/\partial z>0$) curves.

\clearpage

\ \vspace{2cm}

\begin{figure}[h]
\includegraphics[width=14cm]{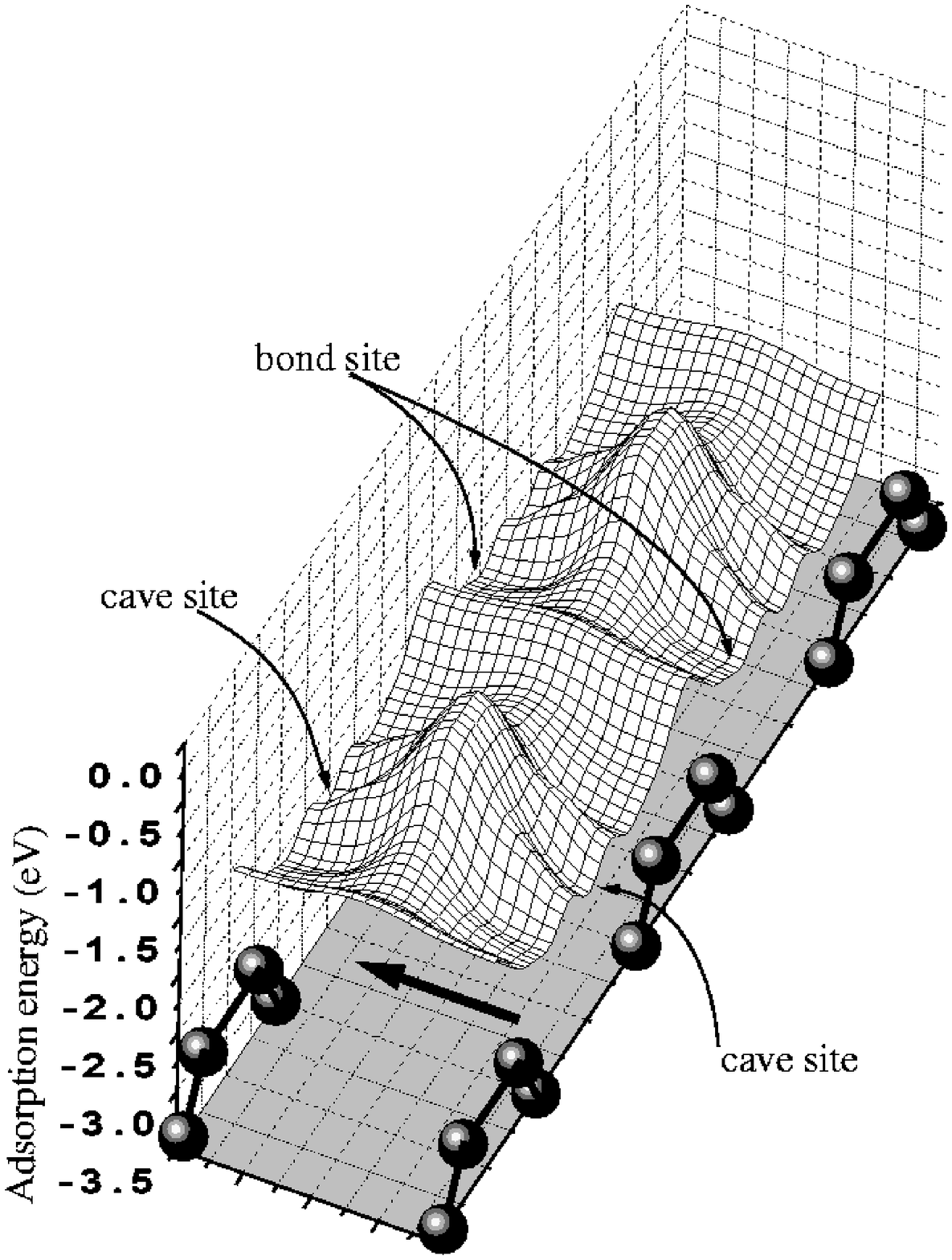}\label{AgSi}
\end{figure}

\vfill\centerline{\bf Figure 1}

\clearpage

\ \vspace{5cm}
\begin{figure}[h]
\includegraphics{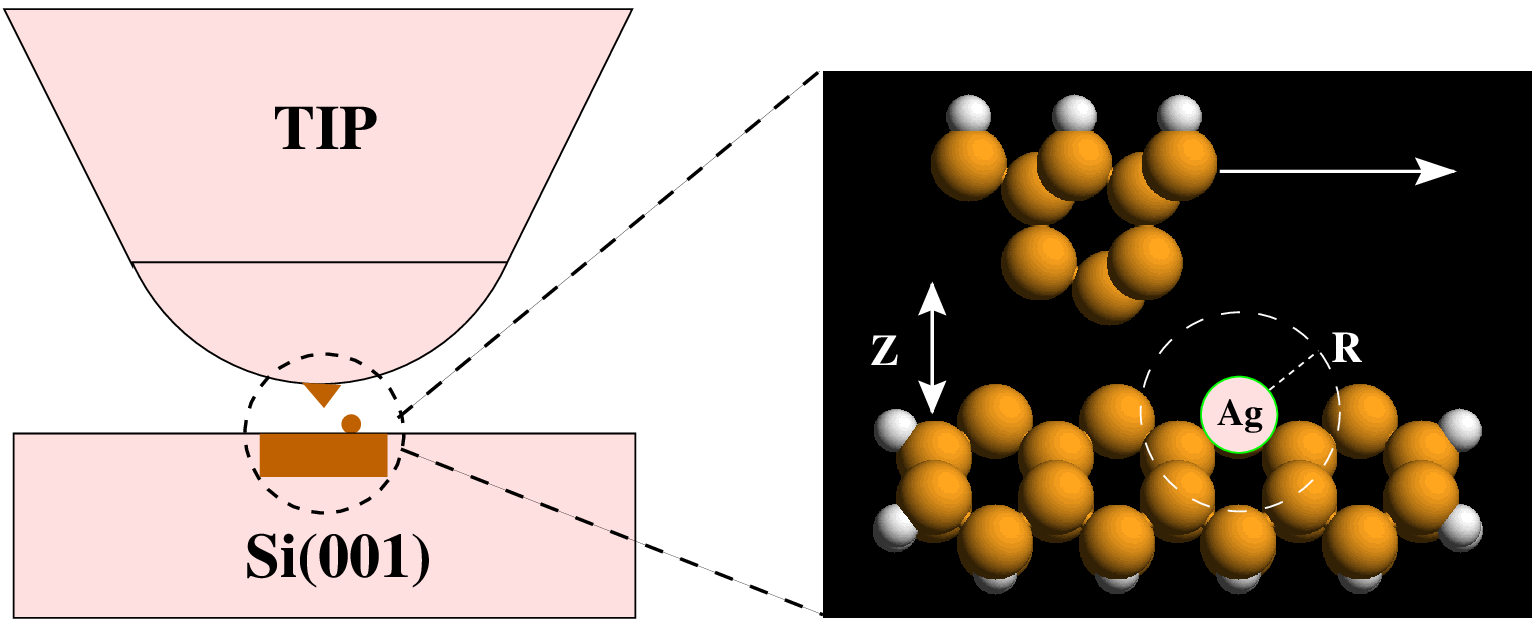}\label{models}
\end{figure}

\vfill\centerline{\bf Figure 2}

\clearpage

\ \vspace{3cm}
\begin{figure}[h]
\includegraphics[width=15cm]{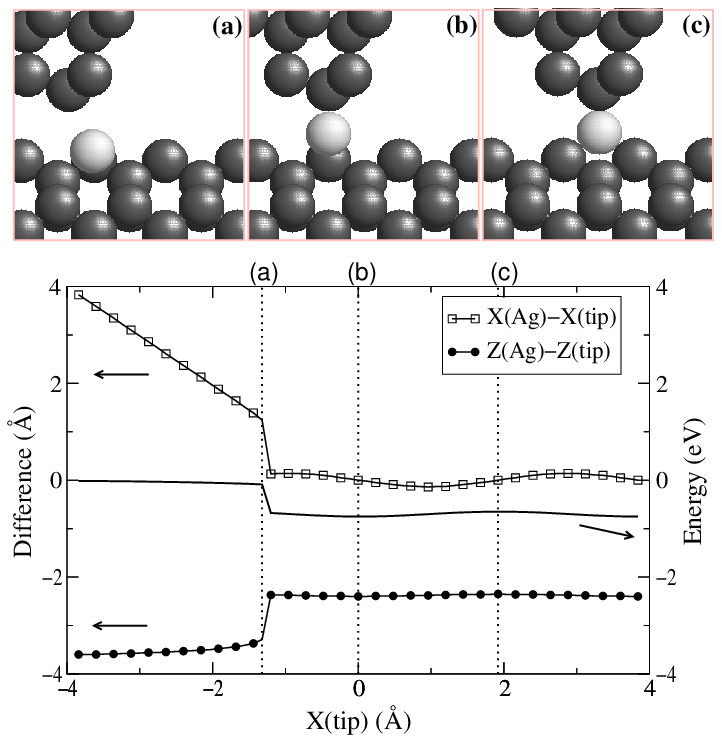} 
\end{figure}

\vfill\centerline{\bf Figure 3}

\clearpage

\ \vspace{3cm}
\begin{figure}[h]
\includegraphics[width=15cm]{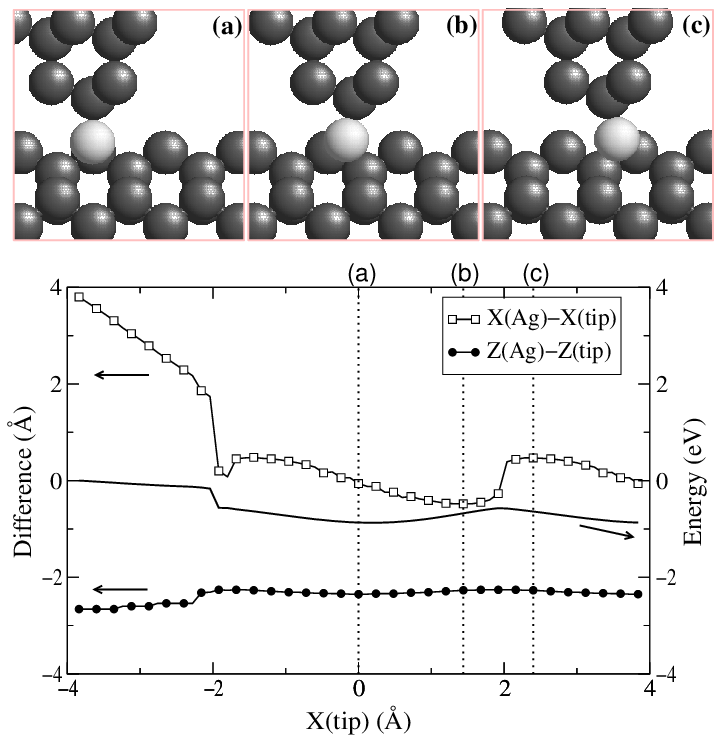}
\end{figure}

\vfill\centerline{\bf Figure 4}

\clearpage

\ \vspace{3cm}
\begin{figure}[h]
\includegraphics[width=15cm]{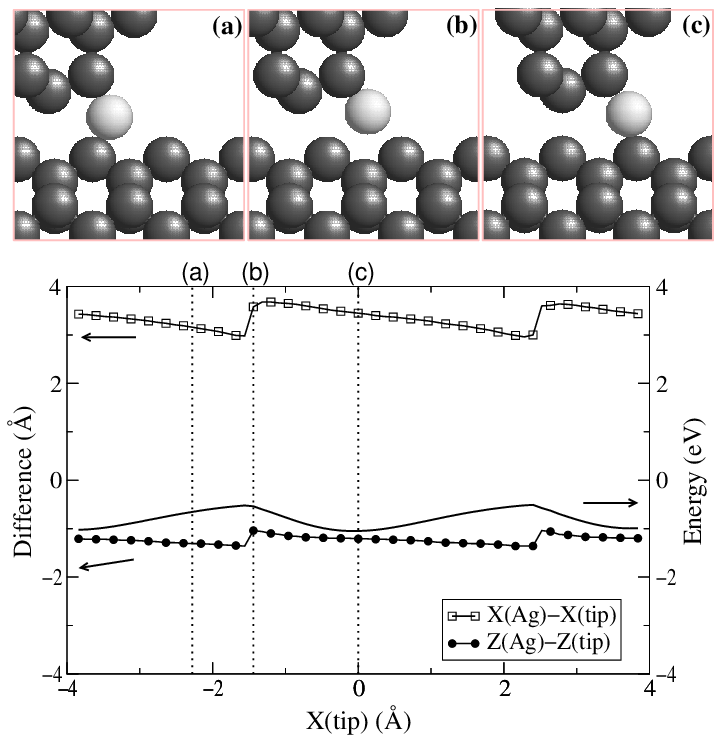}
\end{figure}

\vfill\centerline{\bf Figure 5}

\clearpage

\ \vspace{3cm}
\begin{figure}[h]
\includegraphics[width=15cm]{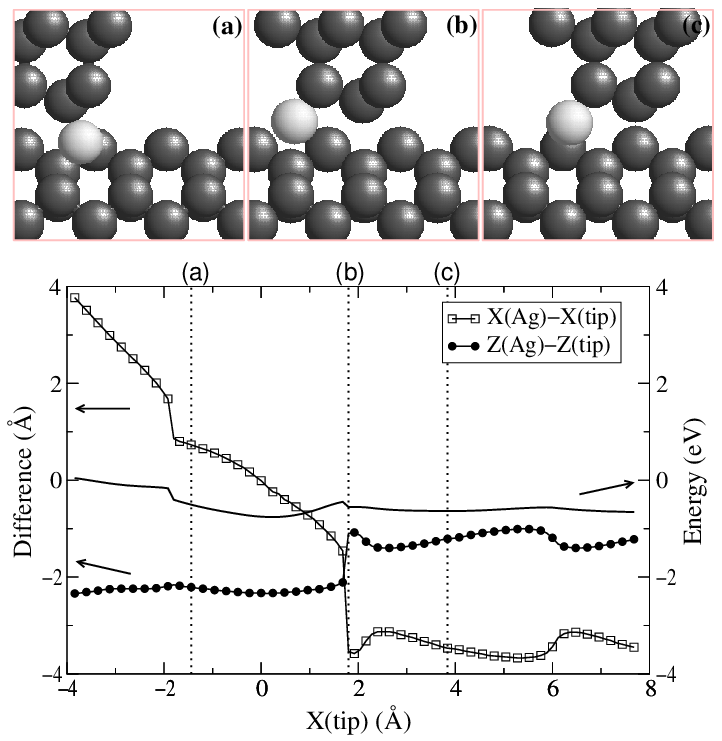}
\end{figure}

\vfill\centerline{\bf Figure 6}

\clearpage

\ \vspace{3cm}
\begin{figure}[h]
\includegraphics[width=17cm]{fig7.eps}
\end{figure}

\vfill\centerline{\bf Figure 7}

\clearpage

\ \vspace{3cm}
\begin{figure}[h]
\includegraphics[width=17cm]{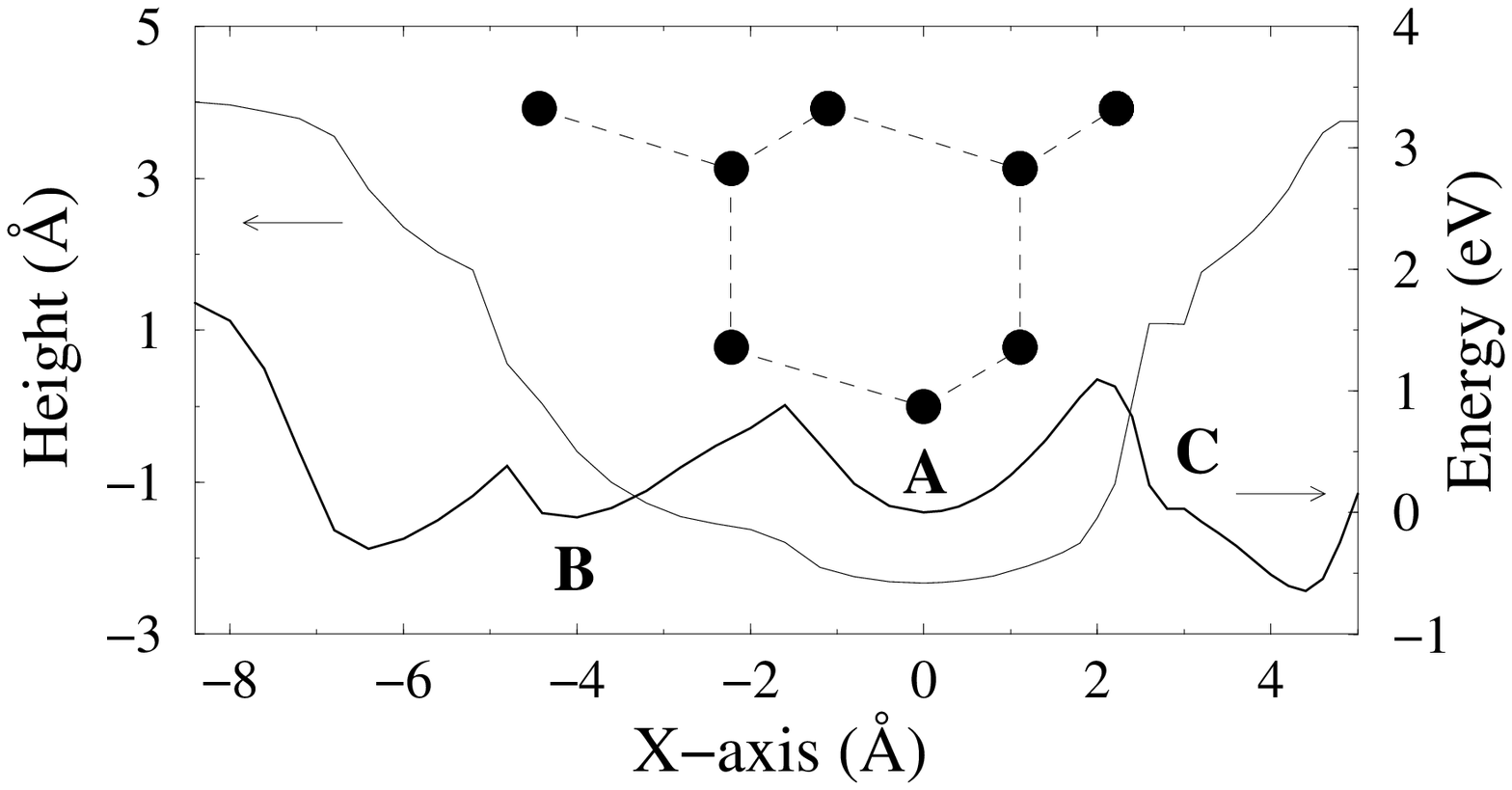}
\end{figure}

\vfill\centerline{\bf Figure 8}

\clearpage

\ \vspace{3cm}
\begin{figure}[h]
\includegraphics[width=17cm]{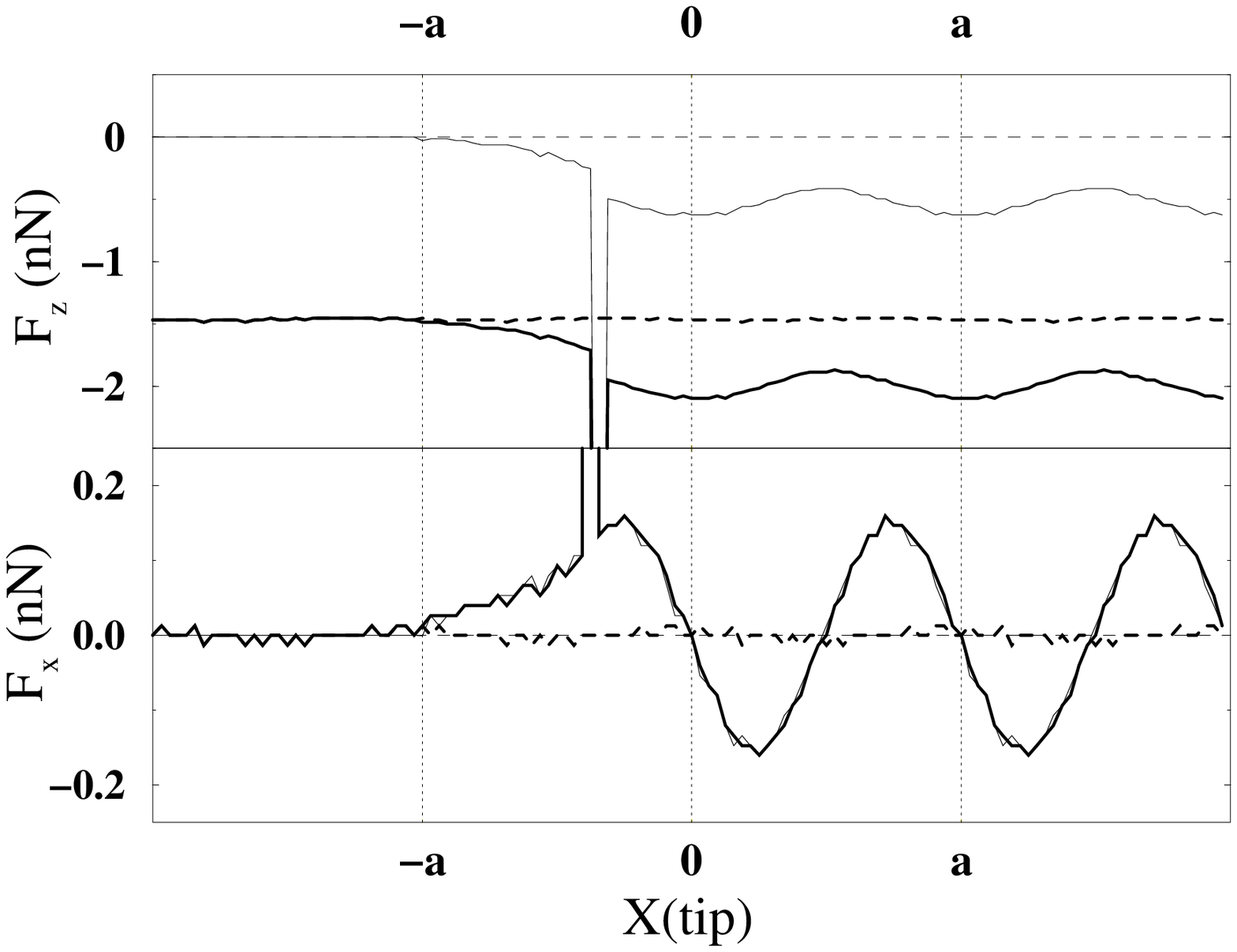}
\end{figure}

\vfill\centerline{\bf Figure 9}

\clearpage

\ \vspace{0.5cm}
\begin{figure}[h]
\includegraphics[width=15cm]{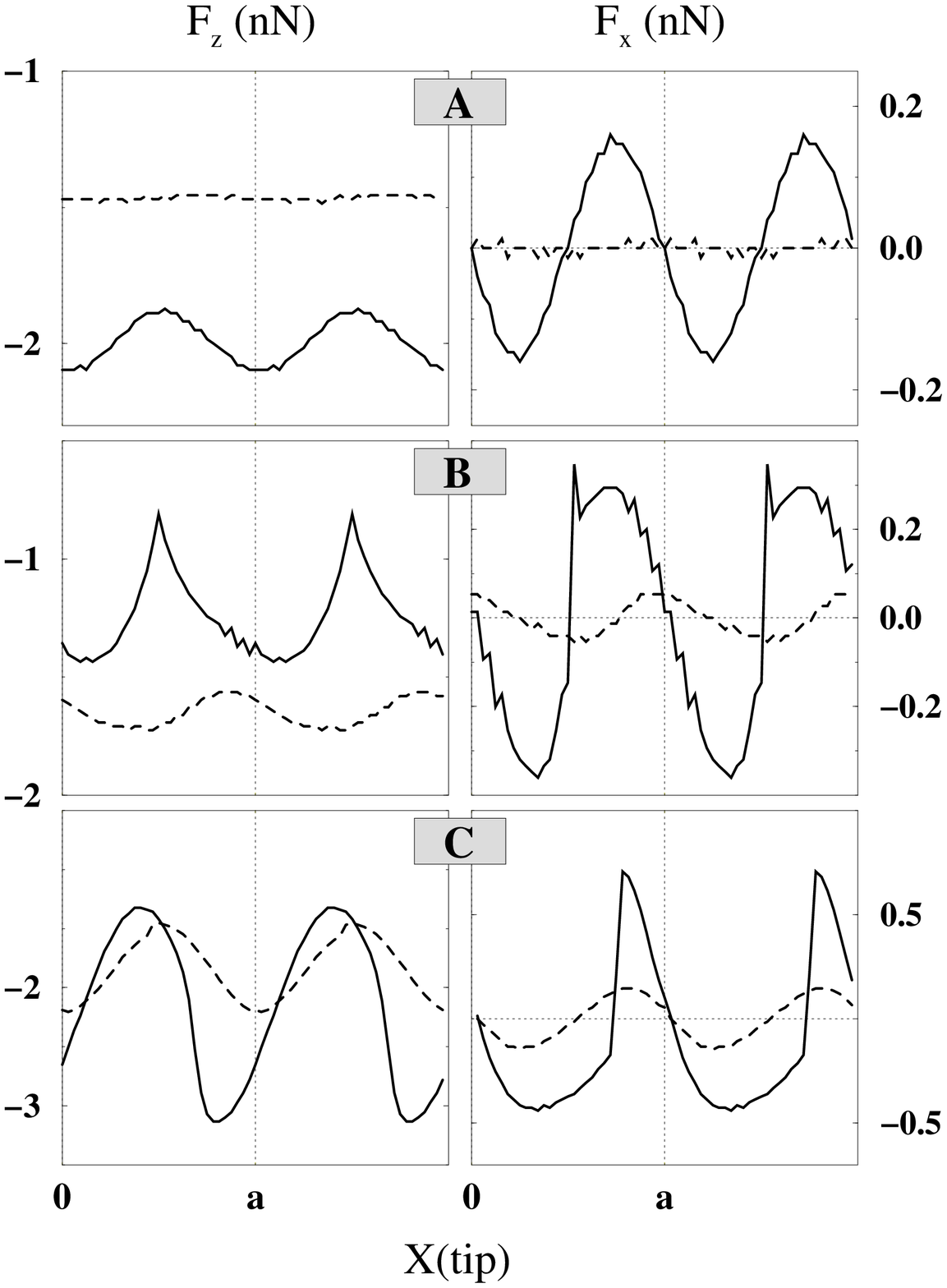}
\end{figure}

\vfill\centerline{\bf Figure 10}

\clearpage

\ \vspace{3cm}
\begin{figure}[h]
\includegraphics[width=17cm]{fig11.eps}
\end{figure}

\vfill\centerline{\bf Figure 11}

\clearpage

\ \vspace{3cm}
\begin{figure}[h]
\includegraphics[angle=-90,width=17cm]{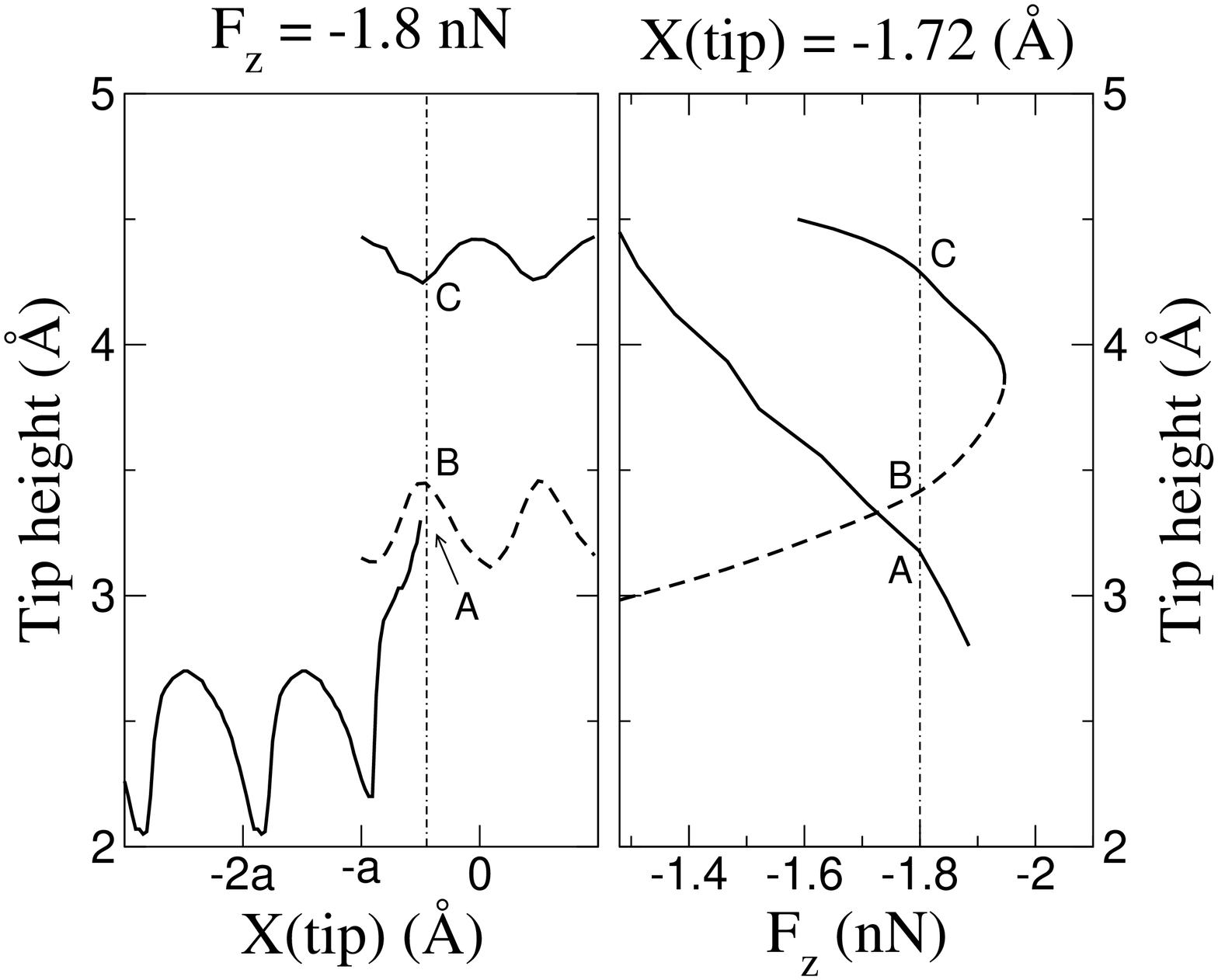}
\end{figure}

\vfill\centerline{\bf Figure 12}

\end{document}